\documentclass[aps,prb,twocolumn,showpacs]{revtex4}
\usepackage{graphicx}
\usepackage{amsmath}
\usepackage{color}

\begin{document}

\title{Exciton-phonon interactions and superconductivity bordering charge order in TiSe$_2$}

\author{Jasper van Wezel}
  \affiliation{Cavendish Laboratory, University of Cambridge, Madingley Road, Cambridge CB3 0HE, UK}
\author{Paul Nahai-Williamson}
  \affiliation{Cavendish Laboratory, University of Cambridge, Madingley Road, Cambridge CB3 0HE, UK}
\author{Siddarth S. Saxena}
  \affiliation{Cavendish Laboratory, University of Cambridge, Madingley Road, Cambridge CB3 0HE, UK}

\begin{abstract}
The strong coupling between lattice modes and charges which leads to the formation of charge density waves in materials such as the transition-metal dichalcogenides may also give rise to superconductivity in the same materials, mediated by the same exciton or phonon modes that dominate the charge ordered state. Such a superconducting phase has recently been observed for example in TiSe$_2$, both upon intercalation with Copper and in the pristine material under pressure. Here we investigate the interplay of exciton formation and electron-phonon coupling within a simplified model description. We find that the combined exciton-phonon modes previously suggested to drive the charge density wave instability in TiSe$_2$ are also responsible for the pairing of electrons in its superconducting regions. Based on these results, it is suggested that both of the observed domes form part of a single superconducting phase. We also study the effect of the quantum critical fluctuations emerging from the suppressed charge order on the transport properties directly above the superconducting region, and compare our finding with reported experimental results.
\end{abstract}


\maketitle

\section{Introduction}

Strongly correlated electron systems have proven to be fertile ground for the realisation of unconventional superconducting states, which often arise upon tuning the electronic order to a quantum phase transition. The critical fluctuations associated with the disappearance of the order parameter at zero temperature influence the physics in a wide range of temperature and tuning parameter values around the quantum critical point, and may be used to directly mediate novel pairing interactions. Non-BCS superconductivity of this type has been found on the borders of antiferromagnetism in heavy fermion materials \cite{Waldram, Mathur} and of charge order in materials such as the nearly ferroelectric SrTiO$_{3}$ and the charge-density-wave (CDW) ordered transition metal dichalcogenides \cite{Schooley, Kusmartseva}.

The CDW phase in some of the latter materials involves exciton formation as well as electron-phonon coupling, thus adding extra interest to the superconducting phase emerging upon its suppression. As was suggested already four decades ago \cite{Little,Ginzburg}, excitons can directly mediate a pairing interaction between electrons, replacing the phonons in the usual BCS construction. It was soon realised that the excitonic `glue' avoids all temperature limits normally imposed by phonons, thus paving the way for room temperature superconductivity \cite{Dolgov}. In practise such an exciton-mediated superconducting phase has turned out to be elusive, because in any real material the renormalisation of the phonon spectrum which typically results from coupling to the excitons tends to cause a structural instability to occur before superconductivity has a chance to arise \cite{Dolgov}. The strong, and seemingly unavoidable, coupling between excitons and phonons makes it impossible for a truly exciton-driven superconducting phase to form. The combined influence of excitons and phonons however may still significantly affect the electronic pairing interactions, and lead to other types of unconventional charge ordered and superconducting phases.

A specific example of a material in which the excitonic and phononic degrees of freedom both contribute significantly to the formation of charge order is given by the CDW phase of TiSe$_2$. The lattice plays an important role in the formation of the CDW in undoped TiSe$_2$ at ambient pressure \cite{Rossnagel, Kidd, Whangbo, Hughes, Wilson78, WilsonIR, Motizuki, Suzuki, UsEPL, UsPRB}, as can be seen for example by the observation in quasi-elastic x-ray scattering experiments of a softening of the $L^{-}_{1}$ phonon mode associated with the CDW distortion \cite{Holt}. At the same time it is clear that excitons also are an indispensable ingredient in the description of this system \cite{UsEPL, UsPRB, Wilson, Zunger, LiCDW, Monney, Cercellier}, as witnessed for example by the large transfer of spectral weight to backfolded bands in ARPES measurements of the CDW phase \cite{Cercellier}. In this context, the recent discoveries of superconductivity in the intercalated compound Cu$_x$TiSe$_2$ by Morosan et al. \cite{Morosan, Morosan2}, and in pure TiSe$_2$ under pressure by Kusmartseva et al. \cite{Anna}, are of great interest.

The superconducting phase which emerges upon suppression of the CDW order is likely to be mediated by a combination of excitons and phonons similar to the one that has been suggested to be responsible for stabilising the charge order \cite{UsPRB}. It is not \textit{a priori} clear however what the effect will be of the involvement of excitons in the electronic pairing mechanism. For example, the superconducting transition temperature could be either suppressed or enhanced. In TiSe$_2$, the CDW can be suppressed either by intercalation with Copper or by applying pressure, and in both cases a superconducting dome emerges which covers the potential quantum critical point. The details of the approach towards criticality however, are very different \cite{Anna}. In the doped compound the Hall coefficient changes sign as Copper is added, suggesting that the TiSe$_2$ layers become doped by electrons. In the pure compound under pressure on the other hand, no such sign change occurs, and the electron density is assumed to remain constant, while the two-band character increases \cite{Klipstein, Friend}. These differences in material properties have previously been suggested to hint at a difference in the nature of the superconducting phases emerging from the suppressed CDW order \cite{Anna,Li,Jishi}. One cause for such discordance could be a change of the role played by the excitons in these compounds.

We recently introduced a generic model system to study the effect of the interplay between exciton formation and electron-phonon coupling on the formation of CDW order \cite{UsEPL,UsPRB,ourselvesDresden}. Within this model it was shown that excitons can compete, coexist or even cooperate with phonons to drive the transition. Here we extent the previous model to also include superconducting pairing interactions (section II). We study the model in two different regimes, and describe the effects of electron doping (section III A) as well as introducing an effective pressure (section III B). We find that although the resulting superconducting phases are BCS-like, the presence of excitons does play an essential role in renormalising the phonon properties. The cooperation of exciton formation and electron-phonon coupling may even lead to a slight enhancement of the superconducting transition temperature. Finally, in section IV, we consider the effect that critical fluctuations emanating from the critical point covered by the superconducting dome may have on the temperature dependence of the resistivity in the normal phase. The critical exponents induced by such fluctuations form an important experimental tool in studying the properties of the hidden quantum critical point \cite{Sachdev,Moriya,Lonzarich,Lonzarich85}, and the nature of the modes driving the instability \cite{Mathur, Saxena, Pines, Smith, Rowley}.

\section{A Model System}

\begin{figure}[t]
\centerline{{\includegraphics[width=0.6 \linewidth]{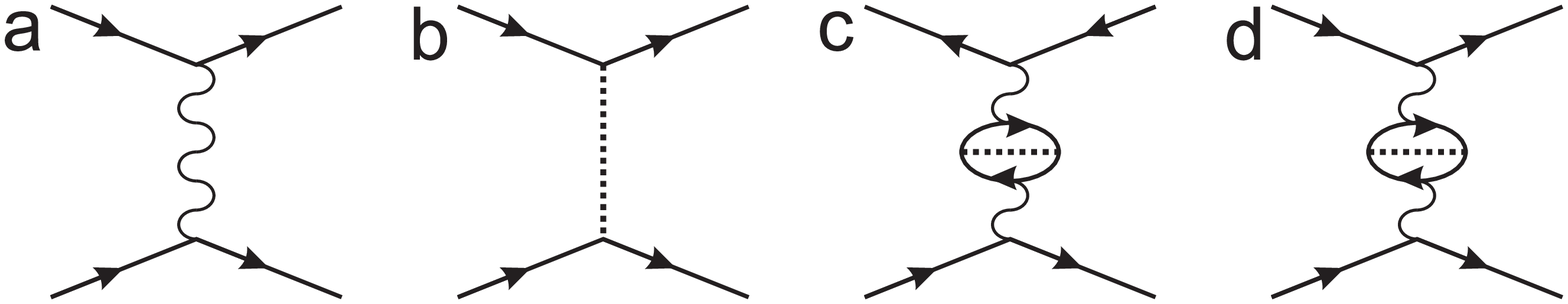}}}
\caption{Feynman diagrams for some of the instabilities inherent in the Hamiltonian of eq.~\eqref{GeneralHamiltonian}. \textbf{(a)} Cooper pair formation via phonon exchange. \textbf{(b)} Cooper pair formation arising from the exciton binding energy. \textbf{(c)} CDW formation promoted by electron-hole interaction via renormalised phonons. \textbf{(d)} Cooper pair formation via renormalised phonon exchange.}
\label{diagrams}	
\end{figure}

The Hamiltonian governing the interplay between excitons, phonons and conduction electrons takes the general form:
\begin{align}
\hat{H} = & \hat{H_0} + \hat{H}_{\text{exc}} + \hat{H}_{\text{e-p}}.
\label{GeneralHamiltonian}
\end{align}
This includes the bands of non-interacting electrons $H_0$, the effective Coulomb interaction between electrons and holes responsible for creating excitons $H_{\text{exc}}$, and the electron-phonon interaction $H_{\text{e-p}}$. A CDW instability may arise either directly from the electron-phonon coupling (such as in a Jahn-Teller driven scenario), directly from the exciton binding energy (forming an excitonic insulator), or from a combination of the two effects (for example via the renormalisation of the phonon spectrum by excitons). Likewise, the standard BCS type of superconductivity arises solely out of the last term in equation \eqref{GeneralHamiltonian}, while a purely exciton-driven superconducting instability is generated by the second term (see figures \ref{diagrams}a and b). The combined effect of both of these excitations, which is thought to underlie the CDW formation in f.e. TiSe$_2$ \cite{UsPRB}, may also lead to superconductivity (figures \ref{diagrams}c and d).

In the following, we will study the possible roles played by excitons and phonons in the emergence of a superconducting phase within a specific realisation of the generic structure of equation \eqref{GeneralHamiltonian} which has previously been suggested to describe the CDW formation in TiSe$_2$ \cite{ourselvesDresden,UsEPL,UsPRB}. The model will be general enough to give a qualitative understanding of the influence of excitons and phonons on the superconductivity observed in materials such as TiSe$_2$, but it will not yield any quantitative results. 

\begin{figure}[t]
\includegraphics[width=0.7\columnwidth]{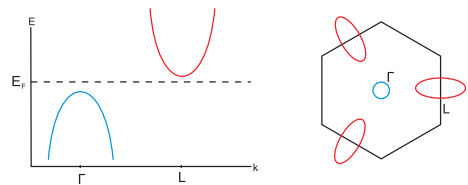}
\caption{\textbf{left} A schematic representation of the band structure very close to the Fermi energy, assuming a small indirect gap between the valence and conduction bands.
\textbf{right} Constant energy maps near E$_F$ forming electron and hole pockets in the first Brillouin zone of TiSe$_2$ (adapted from [\onlinecite{MonneyPLD}]).
}
\label{band}
\end{figure}

\begin{figure}[t]
\includegraphics[width=0.6\columnwidth]{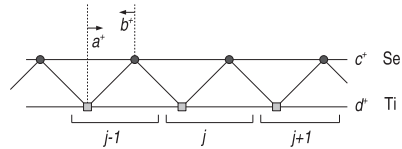}
\caption{The quasi one-dimensional double chain used as a model to study the interplay of electron-phonon coupling and exciton formation. Hoppings between sites on the upper and lower chains are described by combinations of the electronic $c$ and $d$ operators, while lattice deformations are included through the $a$ and $b$ phonon operators.
}
\label{chain}
\end{figure}

To be specific, we only consider the two bands closest to the Fermi energy which are known to be the Se-4$p$ valence band located at the centre of the first Brillouin zone, and the Ti-3$d$ conduction band forming electron pockets at the zone boundaries, as in figure \ref{band}. The Hamiltonian will be defined on the double chain lattice schematically shown in figure \ref{chain}. The $c$ sites represent the Se-4$p$ orbitals, while the $d^{\dagger}$ operators create electrons in the Ti-3$d$ orbitals. 

This greatly simplified quasi one-dimensional model has previously been argued to qualitatively capture the physics of the CDW transition in TiSe$_2$ \cite{ourselvesDresden,UsEPL,UsPRB}. The two-dimensional layers making up the hexagonal TiSe$_2$ structure are separated from each other by a Van der Waal's gap, which allows us to neglect the effects of inter-layer coupling. Moreover, from a tight-binding fit to the band structure it becomes clear that the electronic structure is dominated by a small set of orbital overlap integrals, which indicate that to a first approximation, the system can be thought of as a network of weakly coupled quasi one-dimensional Ti chains and their surrounding Se environment \cite{ourselvesDresden}. Multiple copies of the double chain model of figure \ref{chain} may be put together to form a larger, three-dimensional structure, but because the coupling between different double chains is weak it can be taken to be a higher order effect, and we will neglect it here \cite{UsPRB}.

The non-interacting part of the model can be constructed from the tight binding fit to the band structure and experimentally accessible parameters. Together they form the bare potential and kinetic energy terms:
\begin{align}
\hat{H}_{0} = \hbar \omega & \sum_{i} \left( \hat{a}^{\dagger}_{i}\hat{a}^{\phantom \dagger}_{i}+ \hat{b}^{\dagger}_{i}\hat{b}^{\phantom \dagger}_{i} \right) + \frac{\Delta}{2} \sum_{i}(\hat{d}^{\dagger}_{i}\hat{d}^{\phantom \dagger}_{i} - \hat{c}^{\dagger}_{i}\hat{c}^{\phantom \dagger}_{i}) \notag \\
+ \frac{t}{2} & \sum_{i}(\hat{d}^{\dagger}_{i}\hat{d}^{\phantom \dagger}_{i+1} + \hat{c}^{\dagger}_{i}\hat{c}^{\phantom \dagger}_{i+1} + H.c.) \notag \\ 
+ t' & \sum_{i}(\hat{d}^{\dagger}_{i}\hat{c}^{\phantom \dagger}_{i} + \hat{d}^{\dagger}_{i+1}\hat{c}^{\phantom \dagger}_{i} + H.c.).
\end{align}

The exciton binding energy is approximated as a purely local term which favours electrons and holes to be no further than one lattice spacing apart, while the electron-phonon term, encoding for Jahn-Teller effects, couples the $c$ and $d$ orbitals to the phonon operators $\hat{a}^{\dagger}$ and $\hat{b}^{\dagger}$, and their associated dimensionless distortions, $\hat{X}^a = ( \hat{a}^{\dagger} +  \hat{a} ) / \sqrt{2 m \omega a^2}$ and $\hat{X}^b = ( \hat{b}^{\dagger} +  \hat{b} ) / \sqrt{2 m \omega a^2}$:
\begin{align}
\hat{H}_{\text{exc}} = -V \sum_{i} & \left[ \hat{d}^{\dagger}_i \hat{d}^{\phantom \dagger}_i \left( 1 - \hat{c}^{\dagger}_i \hat{c}^{\phantom \dagger}_i \right) \right. \notag \\
& + \left. \hat{d}^{\dagger}_i \hat{d}^{\phantom \dagger}_i \left( 1 - \hat{c}^{\dagger}_{i-1} \hat{c}^{\phantom \dagger}_{i-1} \right) \right],
\label{EXC}
\end{align}
\begin{align}
\hat{H}_{\text{e-p}} = \alpha \sum_{i} & \left[ \left( \hat{X}^a_i + \hat{X}^b_i \right) \left(\hat{d}^{\dagger}_i \hat{c}^{\phantom \dagger}_i + \hat{c}^{\dagger}_i \hat{d}^{\phantom \dagger}_i \right) \right. \notag \\
& - \left. \left( \hat{X}^a_{i+1} + \hat{X}^b_i \right) \left(\hat{d}^{\dagger}_{i+1} \hat{c}^{\phantom \dagger}_i + \hat{c}^{\dagger}_i \hat{d}^{\phantom \dagger}_{i+1} \right) \right] . 
\label{EP}
\end{align}

These interaction terms can be decoupled on the mean field level by introducing parameters for the average lattice distortions ($u=\langle \hat{X}^a_i+\hat{X}^b_i \rangle$), the electron densities on the different sites ($\rho_c = \langle \hat{c}^{\dagger}_i \hat{c}^{\phantom \dagger}_i \rangle$ and $\rho_d = \langle \hat{d}^{\dagger}_i \hat{d}^{\phantom \dagger}_i \rangle$), and the charge transfer between Ti and Se sites both within and between unit cells ($\tau_{\text{in}} = \langle \hat{c}^{\dagger}_i \hat{d}^{\phantom \dagger}_i \rangle$ and $\tau_{\text{out}} = \langle \hat{c}^{\dagger}_i \hat{d}^{\phantom \dagger}_{i+1} \rangle$). In terms of these, the particle number conserving part of the Hamiltonian reads:
\begin{align}
\hat{H}_{MF}=
& \sum_{k} \left\{ \hat{c}^{\dagger}_{k} \hat{c}^{\phantom \dagger}_{k} \left[ t\cos(ka) - \frac{\Delta}{2} + 2V\rho_{d} -\mu \right] \right. \notag \\
& \phantom{\sum} + \hat{d}^{\dagger}_{k} \hat{d}^{\phantom \dagger}_{k} \left[ t\cos(ka) + \frac{\Delta}{2} + 2V(\rho_{c}-1) - \mu \right] \notag \\
& \phantom{\sum} + \hat{d}^{\dagger}_{k} \hat{c}^{\phantom \dagger}_{k} \left[ t'(1+e^{ika}) + \alpha u (1-e^{ika}) \right. \notag \\
& \phantom{\sum + \hat{d}^{\dagger}_{k} \hat{c}^{\phantom \dagger}_{k}} \left. \left. - V(\tau_{\text{in}} + e^{ika}\tau_{\text{out}}) \right] + \text{H.c.} \right\}
\label{Hcdw}
\end{align}

To also include the superconducting pairing interactions between electrons, we follow the usual BCS prescription by expanding $\hat{H}_{\text{e-p}}$ to second order in the electron-phonon coupling. The expansion terms can be decoupled by introducing superconducting Cooper pair densities of s-wave symmetry both within the unit cell ($\sigma_{\text{in}} = \langle \hat{c}^{\phantom \dagger}_i \hat{c}^{\phantom \dagger}_{i} \rangle$) and between unit cells ($\sigma_{\text{out}} = \langle \hat{c}^{\phantom \dagger}_i \hat{c}^{\phantom \dagger}_{i \pm 1} \rangle$). Notice that in principle also interband pairing terms of the form $\langle \hat{c} \hat{d} \rangle$ could arise both from the electron phonon coupling, and directly from the excitonic binding energy \cite{Dolgov,Varma,Littlewood,Nakai}. In the present model these pairing channels turn out to be repulsive, and we will neglect them from here on. The remaining pairing terms are then:
\begin{align}
\hat{H}_{MF}'=
& -\frac{\alpha^2}{m \omega^2 a^2} \sum_{k} \left\{ \hat{d}^{\dagger}_{k} \hat{d}^{\dagger}_{-k} \left[ (2-cos(ka)) \sigma_{\text{in}}^c - \sigma_{\text{out}}^c \right] \right. \notag \\
& \phantom{\sum} + \hat{c}^{\dagger}_{k} \hat{c}^{\dagger}_{-k} \left. \left[ (2-cos(ka)) \sigma_{\text{in}}^d - \sigma_{\text{out}}^d \right] + \text{H.c.} \right\}.
\label{Hsc}
\end{align}

Within this mean field formulation we can investigate the effect of electron doping by adjusting the value of the chemical potential to yield a higher total electron density. In doing so, we assume that the shape of the band structure around the Fermi energy will not be significantly affected by the presence of interlayer dopants \cite{Anna}. The effect of pressure can be included in the model by scaling the orbital overlap integrals, in order to replicate the experimentally observed behaviour of an increased two-band character at constant total electron density \cite{Klipstein, Friend}.

\section{Towards Quantum Criticality}

\begin{figure}[t]
\centerline{{\includegraphics[width=0.8 \linewidth]{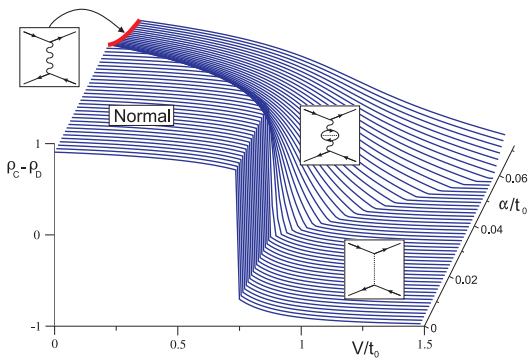}}}
\caption{The mean field solution of equation \eqref{Hcdw} for the charge transfer $\rho_c-\rho_d$. Starting from the undistorted semiconducting phase at the origin, increasing the electron-phonon coupling $\alpha / t$ past a critical value induces a CDW transition, while for any given value of $\alpha$ an increase in the exciton binding energy $V / t$ is seen to enhance the tendency towards CDW formation. At even higher values of $V$, the exciton formation saturates the charge transfer and a uniform excitonic insulator is formed.
}
\label{phases}
\end{figure}

Solving the mean field equations \eqref{Hcdw} self consistently for varying values of the exciton binding energy $V$ and the strength of the electron-hole coupling $\alpha$, we find three distinct phases at zero temperature \cite{UsEPL,UsPRB}. At low $V$ and $\alpha$, the system is in its semiconducting ground state (see figure \ref{phases}). Upon increasing the exciton binding energy a first order transition is encountered at which charges from the occupied valence band transfer to the unoccupied conduction band and form excitonic bound states with the positively charged holes left behind. This state is an (excitonic) insulator, but because we did not include any direct exciton-phonon coupling in the Hamiltonian of equation \eqref{Hcdw}, it does not break the translational symmetry of the lattice. 

Keeping a low exciton binding energy and increasing the electron-phonon coupling instead leads to a second order transition into a CDW ordered phase. Increasing the exciton formation at any given value of the electron-phonon coupling enhances the tendency towards charge order, and indeed there is a region in phase space where a finite exciton binding energy is required in order for a CDW to form. As $V$ is increased even further, an excitonic insulator will once again be formed and translational invariance restored. The CDW observed in pristine TiSe$_2$ is believed to be of a hybrid exciton-phonon character, as in the CDW phase of figure \ref{phases}. Thus excitons as well as phonons play an important role in establishing charge order, and both are expected to play an important role in the superconducting phase that appears as the CDW is suppressed.

\begin{figure}[t]
\centerline{{\includegraphics[width=0.65 \linewidth]{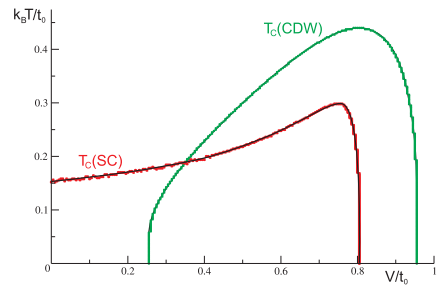}}}
\caption{The transition temperature in the mean field solution of equation \eqref{Hcdw} as a function of the exciton binding energy $V / t$, for a fixed value of the electron-hole coupling $\alpha / t= 0.04$. The transition temperatures of both the superconducting and the charge ordered phase are enhanced by an increase in binding energy due to its renormalisation of the phonon spectrum. At very high values of $V$ the charge transfer is exhausted, and an excitonic insulator is formed instead. Notice that the region of overlap between CDW and superconducting order does not necessarily indicate coexistence.
}
\label{VT}
\end{figure}

If we also include the Cooper pair densities of equation \eqref{Hsc} in the mean field analysis, we find that a superconducting phase may be realised for high enough values of the electron-phonon coupling. The pairing interaction in this phase is mediated by phonons, which in turn are renormalised due to the presence of excitons. If the excitonic binding energy is increased, the renormalisation increases, and the superconducting order parameter, as well as the superconducting T$_C$, may be slightly enhanced, as seen figure \ref{VT}. At even higher values of $V$ there is again an instability towards the formation of an excitonic insulator, and both superconductivity and charge order are lost. Notice that in the present mean field treatment it is not possible to determine whether the CDW and superconducting phases truly coexist in the region of overlap in figure \ref{VT}. In a more detailed theory the superconducting order may well be suppressed by the presence of the CDW phase. Coexistence cannot be directly ruled out however; Kiss et al. show that in conventional charge ordered materials, including the layered transition metal dichalcogenides, CDW and superconducting order can coexist, and that the superconductivity is in fact boosted at points in k-space connected by the CDW ordering vector \cite{Kiss}.

\subsection{Electron Doping}

The CDW observed in pristine TiSe$_2$ under ambient pressure can be suppressed by intercalating with Copper. As the CDW transition temperature is tuned to zero, a superconducting phase has been observed to emerge \cite{Morosan,Morosan2}. To describe this behaviour in the present model of equations \eqref{Hcdw} and \eqref{Hsc}, we assume that the main consequence of intercalation will be the donation of electrons from the Cu intercalants to the Ti conduction band. The addition of Copper is known to also slightly alter the lattice parameters of the TiSe$_2$ layers, acting as an effectively negative pressure \cite{Morosan}. We will assume this to be the smaller effect, and focus instead on the electron doping, which can be modelled as an effective increase in the average total electron density, stabilised by an enhancement of the electronic chemical potential $\mu$.

\begin{figure}[t]
\centerline{{\includegraphics[width=0.6 \linewidth]{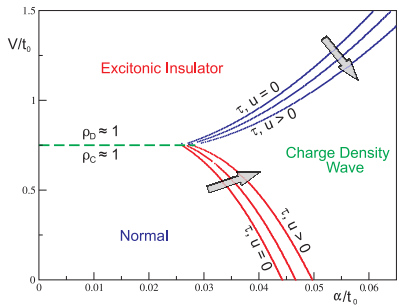}}}
\caption{The mean field phase diagram arising from equations \eqref{Hcdw} and \eqref{Hsc}, for various doping levels. The total charge increases in the direction of the arrows, and takes on the values $\rho_c+\rho_d=1.00$, $1.05$ and $1.10$.
The addition of extra electrons pushes the formation of CDW order to higher values of the electron-phonon coupling, without affecting the onset of the excitonic insulator phase.
}
\label{muphases}
\end{figure}

Upon introducing more electrons into the system described by equation \eqref{Hcdw}, the CDW transition is pushed towards higher values of the electron-phonon coupling, as seen in figure \ref{muphases}. This initial suppression of the charge order can be easily understood in terms of the local physics, by realising that the extra electrons on the $c$ sites block the transfer of charge along the $c$-$d$ bond preferred by the lattice distortions. The transition into the excitonic insulator phase at low values of $\alpha$ is not affected in the same way, because in the absence of a direct exciton-phonon coupling term the formation of excitons happens independently of the lattice deformation.  The order of the transition between normal state and excitonic insulator does change, transforming from a true first order transition at $\rho_c+\rho_d=1$ to a smooth crossover at high dopant levels. 

\begin{figure}[t]
\includegraphics[width=0.27\columnwidth]{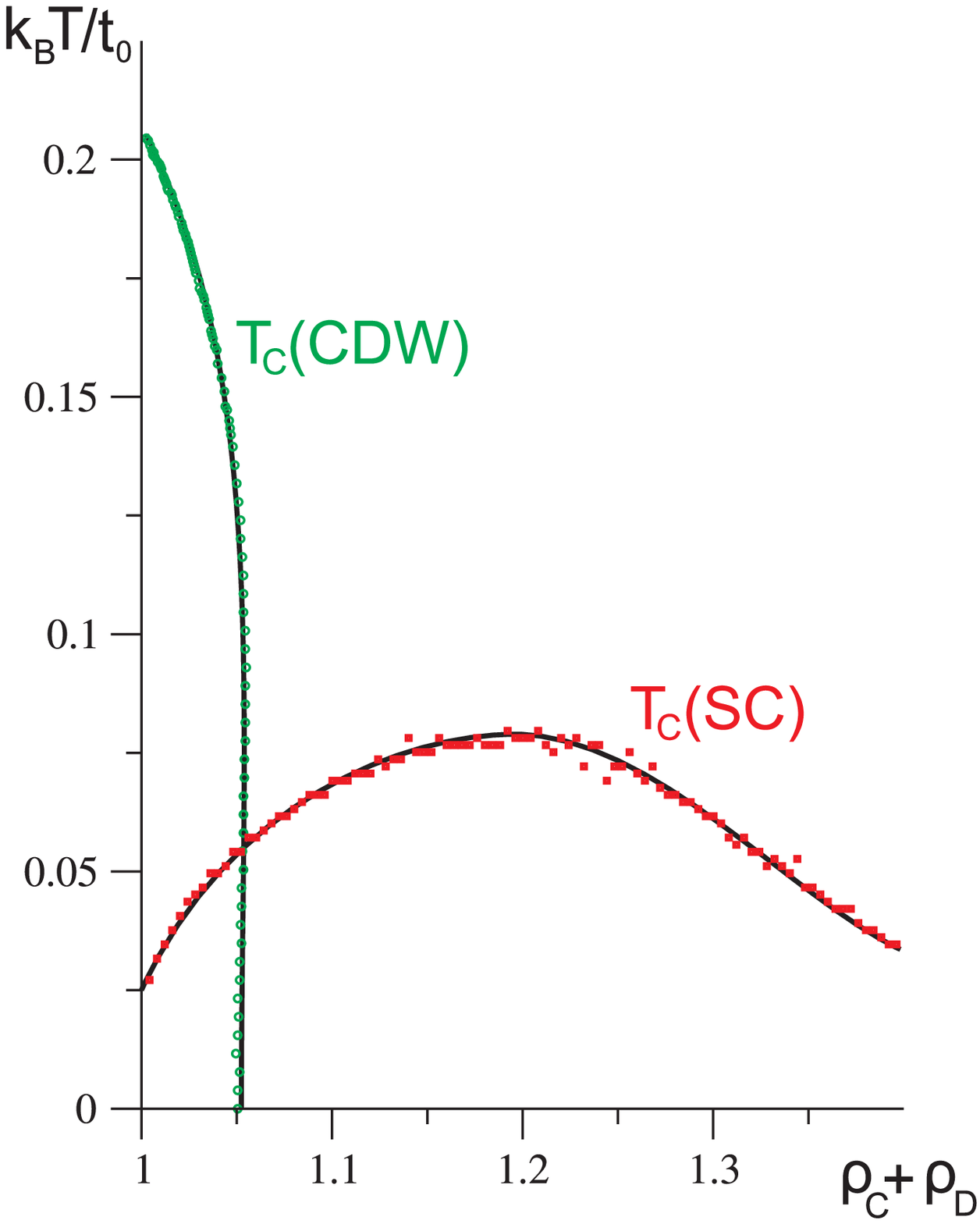} ~~~~~~~ \includegraphics[width=0.5\columnwidth]{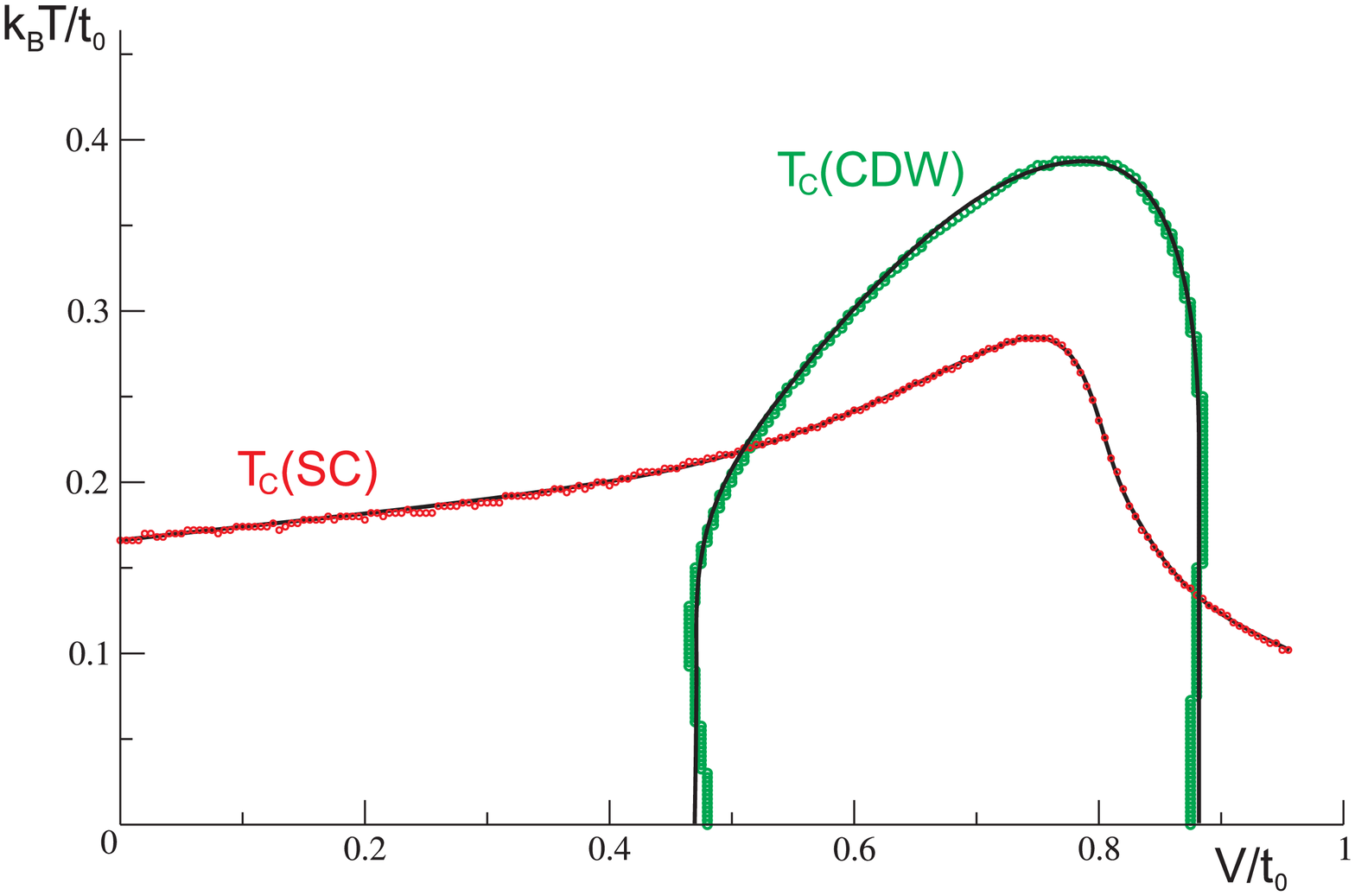}
\caption{The transition temperature in the mean field solution of equations \eqref{Hcdw} and \eqref{Hsc}, in the presence of electron doping: \textbf{left} as a function of the total electron density $\rho_c+\rho_d$, at fixed values for the exciton binding energy ($V/t=0.65$) and the electron-hole coupling ($\alpha / t = 0.032$); and \textbf{right} as a function of the exciton binding energy $V / t$, at a fixed value of the total electron density ($\rho_c+\rho_d=1.1$). \\ As in figure \ref{VT}, the transition temperatures of both the superconducting and the charge ordered phase are enhanced by an increase in excitonic binding energy, indicating the cooperation between the exciton driven renormalisation of the phonon properties and the phonon mediated pairing.
}
\label{domeMu}
\end{figure}

Including also the Cooper pair formation of equation \eqref{Hsc} in the mean field analysis, we find in figure \ref{domeMu} the emergence of a dome shaped superconducting phase surrounding the point where the CDW transition reaches zero temperature. In the present mean field treatment the superconductivity extends to relatively high temperatures, but taking into account fluctuation effects should reduce the extent of the superconducting phase considerably. The positioning of the dome just beyond the point where the disappearing charge order should give rise to a quantum phase transition can be interpreted as indicating that the superconductivity in this region is mediated by the critical fluctuations associated with the parent CDW phase. As before, the pairing ``glue'' thus consists of phonons, renormalised by the presence of excitons. The addition of dopant electrons serves the dual purpose of both directly augmenting the density of carriers available for Cooper pair formation and of driving the system towards the point where the hybrid exciton-phonon modes become soft and can mediate the attractive interaction between the conduction electrons. By plotting the superconducting and charge ordering transition temperatures as a function of the exciton binding energy (see figure \ref{domeMu}), it becomes clear that the excitons play an essential part in this mechanism, strengthening the role of the phonons via the renormalisation of their spectral properties, and enhancing both the CDW order and the superconducting T$_C$. In the present model the charge order is once again cut off at very high values of $V$ due to the formation of the spatially homogeneous excitonic insulator.

\subsection{Pressure}

Superconductivity has also been observed in pure TiSe$_2$ under pressure \cite{Anna}. Based on differences in the behaviour of the Hall coefficient and the sensitivity to magnetic fields it has been argued that the nature of the superconducting phase in this case differs from that of the Copper doped compound. In the model of equations \eqref{Hcdw} and \eqref{Hsc} we can simulate the effects of pressure by assuming that its main influence on the electronic structure arises from a change in the orbital overlap integrals. We thus introduce the rescaled hopping parameters $t=t_0 (1+P_{\text{eff}})$ and $t'=t'_0 (1+c P_{\text{eff}})$ with $c$ a constant of order $1$. Due to the simplified nature of the model we cannot make a quantitative comparison with the experimentally observed behaviour under pressure, but we can use the effective pressure $P_{\text{eff}}$ to qualitatively study its effect on the interplay between exciton formation and electron-phonon coupling.

\begin{figure}[t]
\includegraphics[width=0.6 \columnwidth]{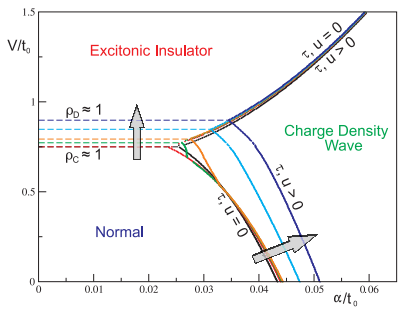} 
\caption{The mean field phase diagram arising from equations \eqref{Hcdw} and \eqref{Hsc}, for various values of the effective pressure $P_{\text{eff}}$. Pressure increases in the direction of the arrows, and takes on the values $P_{\text{eff}}=0.0$, $0.5$, $0.66$, $0.75$, $1.0$ and $1.25$. The application of effective pressure affects both the transition into the CDW phase and the onset of the excitonic insulator phase, pushing them to higher values of the electron-phonon coupling and the exciton binding energy respectively.
}
\label{pressure}
\end{figure}

\begin{figure}[t]
\includegraphics[width=0.4 \columnwidth]{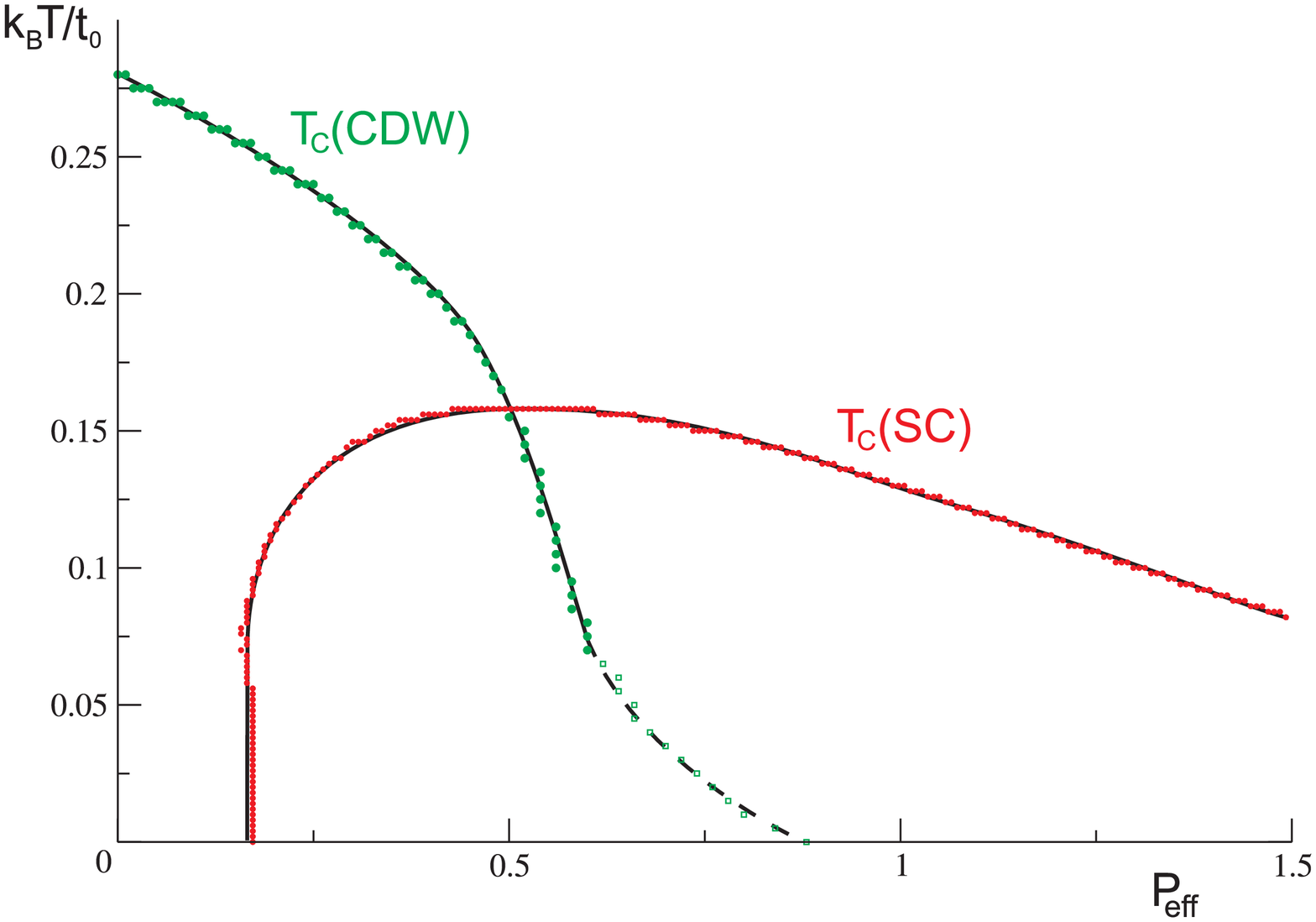} ~~~~~ \includegraphics[width=0.4 \columnwidth]{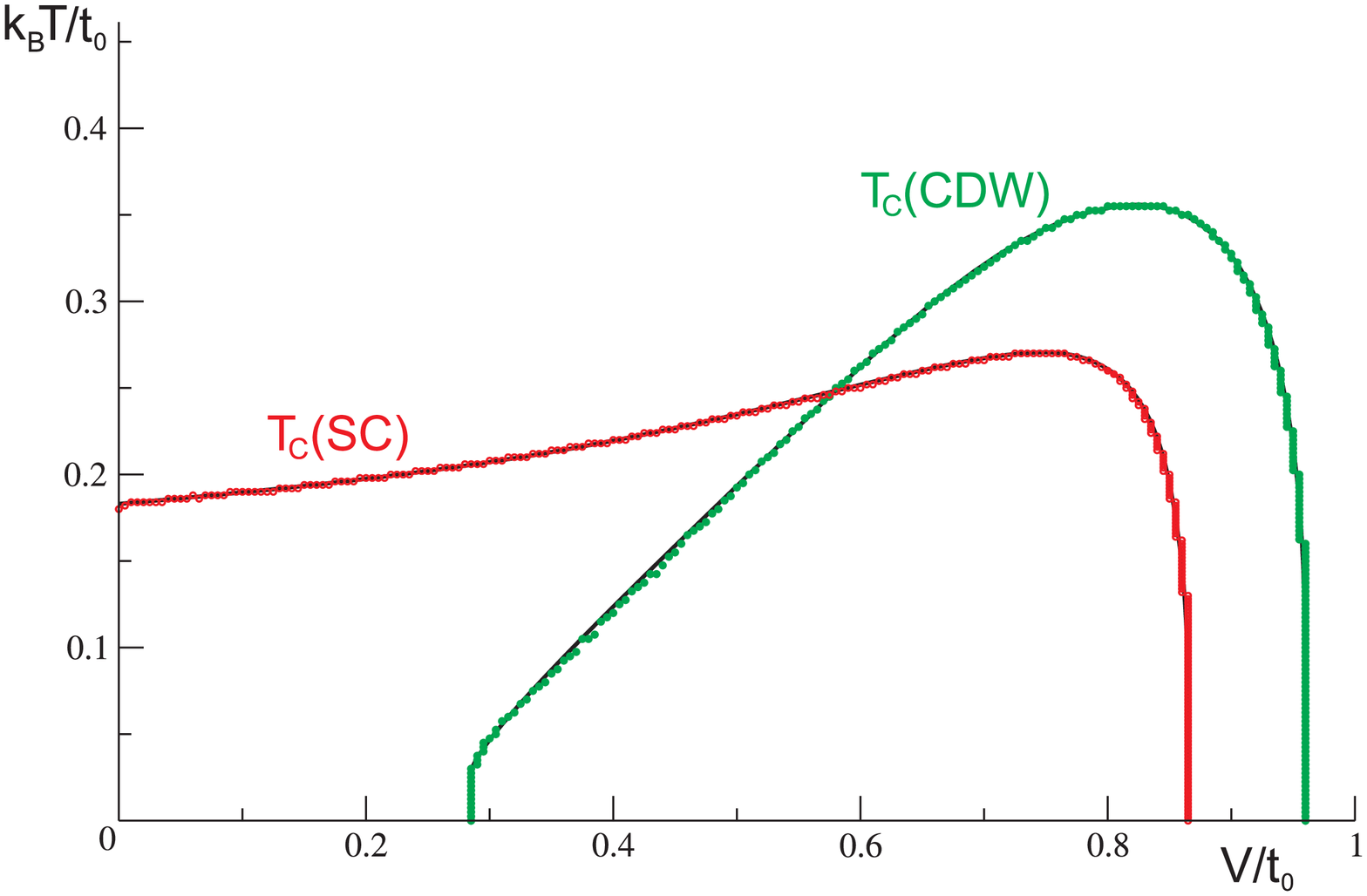}
\caption{The transition temperature in the mean field solution of equations \eqref{Hcdw} and \eqref{Hsc}: \textbf{left} as a function of effective pressure $P_{\text{eff}}$, at fixed values for the exciton binding energy ($V/t_0=0.65$) and the electron-hole coupling ($\alpha / t_0 = 0.035$); and \textbf{right} as a function of the exciton binding energy $V / t_0$, at a fixed value of the effective pressure $P_{\text{eff}} = 0.5$ and $\alpha/ t_0 = 0.04$.
\\ Notice that the CDW transition becomes first order at high values of the effective pressure, and that as in figure \ref{VT}, the transition temperatures of both the superconducting and the charge ordered phase are enhanced by an increase in exciton binding energy, indicating that the cooperation between the exciton driven renormalisation of the phonon properties and the phonon mediated pairing persists under pressure. 
}
\label{dome}
\end{figure}

Figure \ref{pressure} shows the evolution of the CDW phase as a function of effective pressure. It is clear that the excitonic insulating phase and the charge ordered phase are pushed to higher values of the excitonic binding energy and the electron-phonon coupling respectively. At the same time, Cooper pair formation in the regions directly adjacent to the charge ordered phase allow superconductivity to arise as the CDW is suppressed. For any particular set of values for the parameters $\alpha$ and $V$, this implies that the typical behaviour upon tuning both temperature and effective pressure is as shown in figure \ref{dome}. Starting from the CDW phase, a superconducting dome again arises around the point at which the CDW transition reaches zero temperature. As before, the superconducting phase seems to extend to high values of both temperature and effective pressure because of the mean field nature of the present treatment, but is expected to be considerably reduced upon taking into account fluctuation effects.


The pair formation in the superconductor under pressure is again mediated by the same cooperating exciton-phonons that also stabilise the parent CDW phase. The excitonic presence and the associated phonon renormalisation enhances both the charge order at ambient pressure, and the superconducting order at higher pressures, as can be seen in figure \ref{dome}. Even though the mean field treatment given here is expected to largely overestimate the transition temperature, the qualitative trend of increasing T$_C$ as $V$ is increased and the quantum critical point approached, is expected to survive. 

The nature of the critical fluctuations mediating superconductivity in this parameter range is thus no different from that encountered in the electron doped case. In both systems an exciton-enhanced phonon mechanism is the driving force behind both the CDW and the superconducting pairing. The two cases may however vary considerably in the values of their superconducting transition temperatures, which depend both on the pairing mechanism (i.e. the effective coupling strength) and the electronic density of states. The augmented carrier density in the doped compound may thus be expected to support the higher transition temperature. The transport in the normal (high temperature) state is also affected differently in the two regimes: whereas doping explicitly introduces a surplus of electron-like (rather than hole-like) carriers, pressure retains the overall electron-hole balance, which may explain the observed differences in the behaviours of the Hall coefficient. Within the present model this does not affect the pairing mechanism in the two regimes though, and the nature of the critical fluctuations is the same throughout. The superconducting dome in the doped compound can then be continuously connected to the dome in the pure system under pressure, creating a `tunnel' of superconductivity in the doping-pressure-temperature phase diagram, as suggested by Kusmartseva et al. for the case of TiSe$_2$ \cite{Anna}.

\section{Critical Fluctuations}

The exponent of the resistivity in the normal (non-superconducting) state of TiSe$_2$ has been observed to undergo a dip in the pressure range surrounding the critical point at which the CDW order fully disappears \cite{Anna}. Both at higher and lower pressures the low temperature resistivity scales as $\rho \propto T^3$, while at the critical pressure the exponent is seen to be reduced towards $\rho \propto T^{2.5}$. In fact, the exponent may well be reduced even further, as it is not clear from the data presented in reference [\onlinecite{Anna}] whether the reduced exponent has fully converged to its minimal value close to the critical point. Further detailed experiments at low temperatures close to the critical pressure will be required to settle the precise value of the resistivity exponent in this regime. Regardless of its exact depth however, the presence of a dip in the exponent is suggestive of the influence of quantum critical fluctuations, like the ones observed in nearly magnetic materials, heavy fermion compounds and nearly ferroelectric systems \cite{Sachdev,Moriya,Lonzarich,Lonzarich85,Mathur, Saxena, Pines, Smith,Rowley}. Studying these and related critical exponents is an important experimental tool in understanding both the nature of the critical modes and the properties of the quantum critical point.

To calculate the influence of the phononic and excitonic critical modes on the resistivity in our model system, we assume that a description in terms of a Boltzmann transport equation will be applicable. Notice that this implies neglecting any fully quantum mechanical effects, like f.e. weak localisation. Close to the quantum critical point, the collision term in the Boltzmann equation is assumed to be dominated by the scattering of conduction electrons off the bosonic critical modes associated with the (hidden) zero temperature CDW transition. These modes are expected to be the hybrid exciton-phonon excitations discussed in the previous sections. In fact, the main mechanism for scattering conduction electrons, both in the case of excitons and of phonons, consists of the formation of local distortions in the crystal polarisation field. Scattering from such polarisation clouds is well understood in terms of the Fr\"ohlich Hamiltonian \cite{Frohlich1,Frohlich2,Frohlich3}. Although the value of the coupling strength may depend on the extent of the contribution from each mode, the ${\bf k}$-dependence of the resulting matrix elements does not distinguish between excitonic or phononic scatterers. We can thus treat the excitons and phonons on an equal footing, and in the following we will refer to both, as well as to their combination, simply as the bosonic scattering modes.

Following the standard arguments given for example by Ziman \cite{Ziman}, we write the stationary electron distribution as $f_{\bf k} = f^0_{\bf k} - \Phi_{\bf k} \cdot \partial f^0_{\bf k} / \partial \epsilon_{\bf k}$, where $f^0_{\bf k}$ is the Fermi-Dirac distribution and $\Phi_{\bf k}$ is a function to be determined. The resistivity arising from the linearised collision term in the Boltzmann equation is then given by:
\begin{align}
\rho \propto \frac{1}{T} \int d{\bf k} d{\bf k'} ~ \left( \Phi_{\bf k} - \Phi_{\bf k'} \right)^2 P_{{\bf k},{\bf k'}}.
\label{Boltzmann}
\end{align}
Here $P_{{\bf k},{\bf k'}}$ is the scattering rate of electrons with momentum ${\bf k}$ into a state with momentum ${\bf k'}$. Notice that this expression assumes the bosonic scatterers to be in their equilibrium distribution. The function $\Phi_{\bf k}$ is a variational function whose shape is normally well approximated by the ansatz $\Phi_{\bf k} \propto {\bf k} \cdot {\bf E}$, with ${\bf E}$ the applied electric field \cite{Ziman, Dungate, Rice}. Although the lack of isotropy in our model system should in principle be reflected in the variational function, it has been pointed out before that at very low temperatures impurity scattering will typically be strong enough to completely determine the electron distribution, while the resistivity is set by the sum of the terms due to impurity and the critical mode scattering \cite{Ueda}. In this regime the standard ansatz for $\Phi_{\bf k}$ thus remains applicable.

The scattering rate in equation \eqref{Boltzmann} can be related to the dynamic susceptibility using Fermi's golden rule and the fluctuation-dissipation theorem \cite{Ziman,Dungate}:
\begin{align}
P_{{\bf k},{\bf k'}} \propto M_{{\bf k},{\bf k'}}^2 f^0_{\bf k}  \left( 1 - f^0_{\bf k'} \right) & n \left( \epsilon_{\bf k'}-\epsilon_{\bf k} \right) \notag \\
\text{Im} ~ \chi & \left( {\bf k'} - {\bf k},\epsilon_{\bf k'}-\epsilon_{\bf k} \right) .
\label{rate}
\end{align}
Here $n(\omega)$ is the distribution function for the bosonic scattering mode, while $M_{{\bf k},{\bf k'}}$ are the scattering matrix elements. Because both excitons and phonons scatter conduction electrons through a distortion in the local polarisation field, these matrix elements are expected to be well approximated by $M_{{\bf k},{\bf k'}}^2 \propto  \left| {\bf k'} - {\bf k} \right|^2 / \omega_{{\bf k'} - {\bf k}}$, with $\omega$ the dispersion of the bosonic scattering mode \cite{WilsonBook}. 

Away from the quantum critical region, where the susceptibility has no particular features, the expressions of equations \eqref{Boltzmann} and \eqref{rate} are of the same generic form as the corresponding equations in for example the elemental transition metals with their partially occupied $s$ and $d$ bands. In fact, it was pointed out by Wilson that interband transitions in these materials, which necessarily involve finite momentum bosonic modes, will typically give rise to a $\rho \propto T^3$ temperature dependence of the resistivity \cite{Wilson2}. The normal state resistivity observed in TiSe$_2$ far from the transition can thus be straightforwardly interpreted as a consequence of the scattering of conduction electrons between the hole pocket at the centre of the first Brillouin zone and the electron pockets at its boundary.

As the transition is approached, $\chi(q,\omega)$ starts to develop a singularity and the bosonic modes begin to soften, developing a local minimum in their dispersion at the ordering wavevector. Rewriting ${\bf k'} - {\bf k} = {\bf Q} + {\bf q}$, with  $q$ small compared to the ordering wavevector $Q$, we find $\omega \sim a + b q^2$, so that the scattering matrix elements become nearly constant in $q$ in this regime. Further simplification can be achieved by rewriting the momentum integrals as $\int d {\bf k} \propto \int d^2 k \int d \epsilon_k$. After also introducing $\epsilon_{k'}= \omega + \epsilon_{k}$, the integral over $\epsilon_{k}$ can be performed explicitly, and the integrations over constant energy surfaces are simplified to yield \cite{Ueda,Ueda2,WilsonBook}:
\begin{align}
\rho \propto \frac{1}{T} \int d q d \omega ~ q^2 \omega n(\omega) \left( n(\omega)+1 \right) ~ \text{Im} ~ \chi \left( q + Q , \omega \right) .
\label{integral}
\end{align}

Close to the quantum critical region, the bosonic modes become dissipative, and the susceptibility can be approximated in analogy to the overdamped harmonic oscillator \cite{Lonzarich}, 
\begin{align}
\chi(q+Q,\omega)^{-1} = \chi_{q+Q}^{-1} \left( 1 - i \frac{\omega}{\gamma ( \chi^{-1} + c q^2 )} \right).
\label{chi}
\end{align}
 Here we introduced both the static susceptibility $\chi_{q+Q}$ and the uniform susceptibility $\chi$. Notice that the relaxation rates $\gamma$ are nearly momentum independent for scattering processes at the finite wavevector $q+Q$, regardless of the microscopic damping mechanism involved. Inserting this form into equation \eqref{integral}, the energy integral can now be evaluated. Its low temperature expansion is well known \cite{Lonzarich}, and to first order in $T$ the resistivity can be shown to be:
\begin{align}
\rho \propto \int d q ~ q^2 \frac{ \left[ \gamma \left( \chi^{-1} + c q^2 \right) / T \right]^{-1} }{ 1 + 3 / \pi \left[ \gamma \left( \chi^{-1} + c q^2 \right) / T \right] }.
\label{Gil}
\end{align}

Depending on how closely the quantum critical point is approached, this expression can in principle give rise to two different temperature dependencies of the resistivity. In the regime near enough to the transition to render the bosonic modes overdamped, but far enough to ensure that $\chi^{-1}$ is finite and the integrand of equation \eqref{Gil} non-singular, dimensional analysis suffices to show that the resistivity will scale as $\rho \propto T^2$. At the critical pressure on the other hand, the uniform susceptibility becomes identically zero, and the integral is dominated by its low-$q$, singular part. Introducing $x=3 \gamma c q^2 / ( \pi T)$, and cutting off the integration at a temperature independent point $x_c$, we then find \cite{Lonzarich}:
\begin{align}
\rho \propto (T)^{3/2} \int \frac{d x}{\sqrt{x}}.
\label{final}
\end{align}

Whether both of these regimes are in fact realised in TiSe$_2$, and whether they occupy any appreciable portion of phase space, depends largely on the detailed properties of the involved bosonic modes and their couplings to the conduction electrons. As the transition is approached however, the softening of the boson spectrum, the increased damping and the rising singularity in the susceptibility will lead to a renormalised temperature dependence of the resistivity. In the regime precisely above the quantum critical point, the expression of equation \eqref{final} should thus be applicable at very low temperatures. The fact that the experimentally observed temperature dependence does not match the $\rho \propto T^{1.5}$ resistivity found here may be due to the superconducting dome preventing measurements to be made at sufficiently low temperatures. The reported dip in resistivity may then be interpreted as the onset of a crossover from the normal state conductivity towards the full quantum critical behaviour at even lower temperatures.

\section{Conclusions}

The superconducting phase which has been observed in TiSe$_2$ to arise either upon electron-doping the material or by applying hydrostatic pressure is likely to be mediated by the same bosonic modes that are responsible for the charge density wave order in the pristine parent compound at ambient pressure. Because the charge order instability in pure TiSe$_2$ is known to involve both exciton formation and a strong electron-phonon coupling, it may be expected that the superconducting phase will likewise involve a combination of both excitonic and phononic contributions. Within the simplified model discussed in this paper, it has been shown that the presence of excitons in a system already prone to lattice deformation due to a strong electron-phonon coupling may lead to an enhancement of the CDW order. Upon suppression of that order, either by raising the effective chemical potential to inject more electrons or by altering the model parameters to simulate an effective pressure, a superconducting dome appears around the point where the CDW transition reaches zero temperature. In both cases the superconducting order may be enhanced by enhancing the excitonic binding energy, thus revealing the supporting role the excitons play also in this phase. In fact, the cooperation between exciton formation and electron-phonon coupling seems to be ubiquitous throughout the emerging phase diagram. The extent of hybridisation between excitonic and phononic modes may be sensitive to the experimental route taken, but the superconducting Cooper pairs are always held together by some combination of the two modes.

To assess precisely which portion of the superconducting `glue' in TiSe$_2$ consists of excitons, more detailed theoretical modelling (evaluating the full frequency dependent Eliashberg equations for a more realistic model system) as well as new experimental work (focusing specifically on the differences between excitons and phonons) will be required. 
Possible experimental signatures of the influence of excitons may be sought for example by altering the ratio of exciton binding energy to phonon energy $V/\omega$ through the use of isotope effects, or more directly by imposing an altered ratio of exciton binding energy to electron-phonon coupling $V/\alpha$ in stretched thin films on a mismatched substrate. 

The critical fluctuations associated with the quantum critical point below the superconducting dome are experimentally accessible through their influence on the temperature dependence of various thermodynamic quantities. The observed dip in the exponent of the resistivity of TiSe$_2$ under pressure may thus be interpreted as a consequence of the presence of such fluctuations. We have shown in this paper that regardless of the precise origin of the critical fluctuations (whether they are excitons, phonons or a combination thereof), it can be expected on very general grounds that the $\rho \propto T^3$ law which is a consequence of the indirect coupling between two bands in the normal state, will be reduced to a $\rho \propto T^{1.5}$ dependence at very low temperatures close to the critical point because of the critical fluctuations. Depending on the effective damping rates of the modes as the critical region is approached, there may also be an additional region of $\rho \propto T^2$ behaviour between these two extremes. Based on these findings, we suggest that the reported dip in the resistivity exponent may not have reached its absolute minimum yet, and should be interpreted as an indication that critical fluctuations play an important part in the physics of TiSe$_2$ in the investigated pressure range, but that the full quantum critical behaviour will emerge only at lower temperatures. Measurement of related critical exponents, such as those of the specific heat or susceptibility may help to shed further light on this issue.

If the main difference between the two superconducting domes previously observed in TiSe$_2$ is the detailed balance between the effects of exciton formation and electron-phonon coupling, as suggested in this paper, they should be connected by a continuous `tunnel' of superconducting order in the pressure-doping phase diagram \cite{Anna}. Completion of the experimental phase diagram by mapping out the pressure dependence of variously doped compounds, as well as considering other (electron or hole) dopants may thus provide further support for the hybrid exciton-phonon picture presented here. The maximum critical temperature of this superconducting phase will be enhanced in areas where the exciton binding energy effectively cooperates with the phonons, and where a large electronic density of states is available to utilise these cooperating bosonic modes. The amount of enhancement however, will be sensitive to the detailed material properties, and a substantial increase of T$_C$ through this mechanism will likely only be achievable with intricate fine tuning of all involved parameters. 

~\\ \noindent
{\textbf{Acknowledgements}}
~\\ \noindent
The authors would like to thank the EPSRC and Jesus and Homerton Colleges of the University of Cambridge for financial support.


\begin{thebibliography}{55}
\expandafter\ifx\csname natexlab\endcsname\relax\def\natexlab#1{#1}\fi
\expandafter\ifx\csname bibnamefont\endcsname\relax
  \def\bibnamefont#1{#1}\fi
\expandafter\ifx\csname bibfnamefont\endcsname\relax
  \def\bibfnamefont#1{#1}\fi
\expandafter\ifx\csname citenamefont\endcsname\relax
  \def\citenamefont#1{#1}\fi
\expandafter\ifx\csname url\endcsname\relax
  \def\url#1{\texttt{#1}}\fi
\expandafter\ifx\csname urlprefix\endcsname\relax\def\urlprefix{URL }\fi
\providecommand{\bibinfo}[2]{#2}
\providecommand{\eprint}[2][]{\url{#2}}

\bibitem[{\citenamefont{Waldram}(1996)}]{Waldram}
\bibinfo{author}{\bibfnamefont{J.~R.} \bibnamefont{Waldram}},
  \emph{\bibinfo{title}{Superconductivity of Metals and Cuprates}}
  (\bibinfo{publisher}{Taylor \& Francis}, \bibinfo{year}{1996}).

\bibitem[{\citenamefont{Mathur et~al.}(1998)\citenamefont{Mathur, Grosche,
  Julian, Walker, Freye, Haselwimmer, and Lonzarich}}]{Mathur}
\bibinfo{author}{\bibfnamefont{N.~D.} \bibnamefont{Mathur}},
  \bibinfo{author}{\bibfnamefont{F.~M.} \bibnamefont{Grosche}},
  \bibinfo{author}{\bibfnamefont{S.~R.} \bibnamefont{Julian}},
  \bibinfo{author}{\bibfnamefont{I.~R.} \bibnamefont{Walker}},
  \bibinfo{author}{\bibfnamefont{D.~M.} \bibnamefont{Freye}},
  \bibinfo{author}{\bibfnamefont{R.~K.~W.} \bibnamefont{Haselwimmer}},
  \bibnamefont{and} \bibinfo{author}{\bibfnamefont{G.~G.}
  \bibnamefont{Lonzarich}}, \bibinfo{journal}{Nature}
  \textbf{\bibinfo{volume}{394}}, \bibinfo{pages}{39} (\bibinfo{year}{1998}).

\bibitem[{\citenamefont{Schooley et~al.}(1965)\citenamefont{Schooley, Hosler,
  Ambler, Becker, Cohen, and Koonce}}]{Schooley}
\bibinfo{author}{\bibfnamefont{J.~F.} \bibnamefont{Schooley}},
  \bibinfo{author}{\bibfnamefont{W.~R.} \bibnamefont{Hosler}},
  \bibinfo{author}{\bibfnamefont{E.}~\bibnamefont{Ambler}},
  \bibinfo{author}{\bibfnamefont{J.~H.} \bibnamefont{Becker}},
  \bibinfo{author}{\bibfnamefont{M.~L.} \bibnamefont{Cohen}}, \bibnamefont{and}
  \bibinfo{author}{\bibfnamefont{C.~S.} \bibnamefont{Koonce}},
  \bibinfo{journal}{Phys. Rev. Lett.} \textbf{\bibinfo{volume}{14}},
  \bibinfo{pages}{305} (\bibinfo{year}{1965}).

\bibitem[{\citenamefont{Sipos et~al.}(2008)\citenamefont{Sipos, Kusmartseva,
  Akrap, Berger, Forr{\'o}, and Tutis}}]{Kusmartseva}
\bibinfo{author}{\bibfnamefont{B.}~\bibnamefont{Sipos}},
  \bibinfo{author}{\bibfnamefont{A.~F.} \bibnamefont{Kusmartseva}},
  \bibinfo{author}{\bibfnamefont{A.}~\bibnamefont{Akrap}},
  \bibinfo{author}{\bibfnamefont{H.}~\bibnamefont{Berger}},
  \bibinfo{author}{\bibfnamefont{L.}~\bibnamefont{Forr{\'o}}},
  \bibnamefont{and} \bibinfo{author}{\bibfnamefont{E.}~\bibnamefont{Tutis}},
  \bibinfo{journal}{Nat. Mater.} \textbf{\bibinfo{volume}{7}},
  \bibinfo{pages}{960} (\bibinfo{year}{2008}).

\bibitem[{\citenamefont{Little}(1964)}]{Little}
\bibinfo{author}{\bibfnamefont{W.~A.} \bibnamefont{Little}},
  \bibinfo{journal}{Phys. Rev.} \textbf{\bibinfo{volume}{134}},
  \bibinfo{pages}{A1416} (\bibinfo{year}{1964}).

\bibitem[{\citenamefont{Ginzburg}(1965)}]{Ginzburg}
\bibinfo{author}{\bibfnamefont{V.~L.} \bibnamefont{Ginzburg}},
  \bibinfo{journal}{Sov. Phys. JETP} \textbf{\bibinfo{volume}{20}},
  \bibinfo{pages}{1549} (\bibinfo{year}{1965}).

\bibitem[{\citenamefont{Dolgov and Maksimov}(1982)}]{Dolgov}
\bibinfo{author}{\bibfnamefont{O.~V.} \bibnamefont{Dolgov}} \bibnamefont{and}
  \bibinfo{author}{\bibfnamefont{E.~G.} \bibnamefont{Maksimov}},
  \bibinfo{journal}{Sov. Phys. Usp.} \textbf{\bibinfo{volume}{25}},
  \bibinfo{pages}{688} (\bibinfo{year}{1982}).

\bibitem[{\citenamefont{Rossnagel et~al.}(2002)\citenamefont{Rossnagel, Kipp,
  and Skibowski}}]{Rossnagel}
\bibinfo{author}{\bibfnamefont{K.}~\bibnamefont{Rossnagel}},
  \bibinfo{author}{\bibfnamefont{L.}~\bibnamefont{Kipp}}, \bibnamefont{and}
  \bibinfo{author}{\bibfnamefont{M.}~\bibnamefont{Skibowski}},
  \bibinfo{journal}{Phys. Rev. B} \textbf{\bibinfo{volume}{65}},
  \bibinfo{pages}{235101} (\bibinfo{year}{2002}).

\bibitem[{\citenamefont{Kidd et~al.}(2002)\citenamefont{Kidd, Miller, Chou, and
  {T.-C.~Chiang}}}]{Kidd}
\bibinfo{author}{\bibfnamefont{T.~E.} \bibnamefont{Kidd}},
  \bibinfo{author}{\bibfnamefont{T.}~\bibnamefont{Miller}},
  \bibinfo{author}{\bibfnamefont{M.~Y.} \bibnamefont{Chou}}, \bibnamefont{and}
  \bibinfo{author}{\bibnamefont{{T.-C.~Chiang}}}, \bibinfo{journal}{Phys. Rev.
  Lett.} \textbf{\bibinfo{volume}{88}}, \bibinfo{pages}{226402}
  (\bibinfo{year}{2002}).

\bibitem[{\citenamefont{Whangbo and Canadell}(1992)}]{Whangbo}
\bibinfo{author}{\bibfnamefont{M.~H.} \bibnamefont{Whangbo}} \bibnamefont{and}
  \bibinfo{author}{\bibfnamefont{E.}~\bibnamefont{Canadell}},
  \bibinfo{journal}{J. Am. Chem. Soc.} \textbf{\bibinfo{volume}{114}},
  \bibinfo{pages}{9587} (\bibinfo{year}{1992}).

\bibitem[{\citenamefont{Hughes}(1977)}]{Hughes}
\bibinfo{author}{\bibfnamefont{H.~P.} \bibnamefont{Hughes}},
  \bibinfo{journal}{J. Phys. C: Solid State Phys.}
  \textbf{\bibinfo{volume}{10}}, \bibinfo{pages}{L319} (\bibinfo{year}{1977}).

\bibitem[{\citenamefont{Wilson}(1978)}]{Wilson78}
\bibinfo{author}{\bibfnamefont{J.~A.} \bibnamefont{Wilson}},
  \bibinfo{journal}{Phys. Stat. Sol. (b)} \textbf{\bibinfo{volume}{86}},
  \bibinfo{pages}{11} (\bibinfo{year}{1978}).

\bibitem[{\citenamefont{Wilson et~al.}(1978)\citenamefont{Wilson, Jr., Salvo,
  and Ditzenberger}}]{WilsonIR}
\bibinfo{author}{\bibfnamefont{J.~A.} \bibnamefont{Wilson}},
  \bibinfo{author}{\bibfnamefont{A.~S.~B.} \bibnamefont{Jr.}},
  \bibinfo{author}{\bibfnamefont{F.~J.~D.} \bibnamefont{Salvo}},
  \bibnamefont{and} \bibinfo{author}{\bibfnamefont{J.~A.}
  \bibnamefont{Ditzenberger}}, \bibinfo{journal}{Phys. Rev. B.}
  \textbf{\bibinfo{volume}{18}}, \bibinfo{pages}{2866} (\bibinfo{year}{1978}).

\bibitem[{\citenamefont{Motizuki et~al.}(1981)\citenamefont{Motizuki, Suzuki,
  Yoshida, and Takaoka}}]{Motizuki}
\bibinfo{author}{\bibfnamefont{K.}~\bibnamefont{Motizuki}},
  \bibinfo{author}{\bibfnamefont{N.}~\bibnamefont{Suzuki}},
  \bibinfo{author}{\bibfnamefont{Y.}~\bibnamefont{Yoshida}}, \bibnamefont{and}
  \bibinfo{author}{\bibfnamefont{Y.}~\bibnamefont{Takaoka}},
  \bibinfo{journal}{Solid State Comm.} \textbf{\bibinfo{volume}{40}},
  \bibinfo{pages}{995} (\bibinfo{year}{1981}).

\bibitem[{\citenamefont{Suzuki et~al.}(1985)\citenamefont{Suzuki, Yamamoto, and
  Motizuki}}]{Suzuki}
\bibinfo{author}{\bibfnamefont{N.}~\bibnamefont{Suzuki}},
  \bibinfo{author}{\bibfnamefont{A.}~\bibnamefont{Yamamoto}}, \bibnamefont{and}
  \bibinfo{author}{\bibfnamefont{K.}~\bibnamefont{Motizuki}},
  \bibinfo{journal}{J. Phys. Soc. Jpn.} \textbf{\bibinfo{volume}{54}},
  \bibinfo{pages}{4668} (\bibinfo{year}{1985}).

\bibitem[{\citenamefont{van Wezel et~al.}(2010{\natexlab{a}})\citenamefont{van
  Wezel, Nahai-Williamson, and Saxena}}]{UsEPL}
\bibinfo{author}{\bibfnamefont{J.}~\bibnamefont{van Wezel}},
  \bibinfo{author}{\bibfnamefont{P.}~\bibnamefont{Nahai-Williamson}},
  \bibnamefont{and} \bibinfo{author}{\bibfnamefont{S.~S.}
  \bibnamefont{Saxena}}, \bibinfo{journal}{EuroPhys. Letters}
  \textbf{\bibinfo{volume}{89}}, \bibinfo{pages}{47004}
  (\bibinfo{year}{2010}{\natexlab{a}}).

\bibitem[{\citenamefont{van Wezel et~al.}(2010{\natexlab{b}})\citenamefont{van
  Wezel, Nahai-Williamson, and Saxena}}]{UsPRB}
\bibinfo{author}{\bibfnamefont{J.}~\bibnamefont{van Wezel}},
  \bibinfo{author}{\bibfnamefont{P.}~\bibnamefont{Nahai-Williamson}},
  \bibnamefont{and} \bibinfo{author}{\bibfnamefont{S.~S.}
  \bibnamefont{Saxena}}, \bibinfo{journal}{Phys. Rev. B}
  \textbf{\bibinfo{volume}{81}}, \bibinfo{pages}{165109}
  (\bibinfo{year}{2010}{\natexlab{b}}).

\bibitem[{\citenamefont{Holt et~al.}(2001)\citenamefont{Holt, Zschack, Hong,
  Chou, and {T.-C.~Chiang}}}]{Holt}
\bibinfo{author}{\bibfnamefont{M.}~\bibnamefont{Holt}},
  \bibinfo{author}{\bibfnamefont{P.}~\bibnamefont{Zschack}},
  \bibinfo{author}{\bibfnamefont{H.}~\bibnamefont{Hong}},
  \bibinfo{author}{\bibfnamefont{M.~Y.} \bibnamefont{Chou}}, \bibnamefont{and}
  \bibinfo{author}{\bibnamefont{{T.-C.~Chiang}}}, \bibinfo{journal}{Phys. Rev.
  Lett.} \textbf{\bibinfo{volume}{86}}, \bibinfo{pages}{3799}
  (\bibinfo{year}{2001}).

\bibitem[{\citenamefont{Wilson}(1977)}]{Wilson}
\bibinfo{author}{\bibfnamefont{J.~A.} \bibnamefont{Wilson}},
  \bibinfo{journal}{Solid State Comm.} \textbf{\bibinfo{volume}{22}},
  \bibinfo{pages}{551} (\bibinfo{year}{1977}).

\bibitem[{\citenamefont{Zunger and Freeman}(1978)}]{Zunger}
\bibinfo{author}{\bibfnamefont{A.}~\bibnamefont{Zunger}} \bibnamefont{and}
  \bibinfo{author}{\bibfnamefont{A.~J.} \bibnamefont{Freeman}},
  \bibinfo{journal}{Phys. Rev. B} \textbf{\bibinfo{volume}{17}},
  \bibinfo{pages}{1839} (\bibinfo{year}{1978}).

\bibitem[{\citenamefont{Li et~al.}(2007{\natexlab{a}})\citenamefont{Li, Hu,
  Qian, Hsieh, Hasan, Morosan, Cava, and Wang}}]{LiCDW}
\bibinfo{author}{\bibfnamefont{G.}~\bibnamefont{Li}},
  \bibinfo{author}{\bibfnamefont{W.~Z.} \bibnamefont{Hu}},
  \bibinfo{author}{\bibfnamefont{D.}~\bibnamefont{Qian}},
  \bibinfo{author}{\bibfnamefont{D.}~\bibnamefont{Hsieh}},
  \bibinfo{author}{\bibfnamefont{M.~Z.} \bibnamefont{Hasan}},
  \bibinfo{author}{\bibfnamefont{E.}~\bibnamefont{Morosan}},
  \bibinfo{author}{\bibfnamefont{R.~J.} \bibnamefont{Cava}}, \bibnamefont{and}
  \bibinfo{author}{\bibfnamefont{N.~L.} \bibnamefont{Wang}},
  \bibinfo{journal}{Phys. Rev. Lett.} \textbf{\bibinfo{volume}{99}},
  \bibinfo{pages}{027404} (\bibinfo{year}{2007}{\natexlab{a}}).

\bibitem[{\citenamefont{Monney et~al.}(2009{\natexlab{a}})\citenamefont{Monney,
  Cercellier, Clerc, Battaglia, Schwier, Didiot, Garnier, Beck, Aebi, Berger
  et~al.}}]{Monney}
\bibinfo{author}{\bibfnamefont{C.}~\bibnamefont{Monney}},
  \bibinfo{author}{\bibfnamefont{H.}~\bibnamefont{Cercellier}},
  \bibinfo{author}{\bibfnamefont{F.}~\bibnamefont{Clerc}},
  \bibinfo{author}{\bibfnamefont{C.}~\bibnamefont{Battaglia}},
  \bibinfo{author}{\bibfnamefont{E.~F.} \bibnamefont{Schwier}},
  \bibinfo{author}{\bibfnamefont{C.}~\bibnamefont{Didiot}},
  \bibinfo{author}{\bibfnamefont{M.~G.} \bibnamefont{Garnier}},
  \bibinfo{author}{\bibfnamefont{H.}~\bibnamefont{Beck}},
  \bibinfo{author}{\bibfnamefont{P.}~\bibnamefont{Aebi}},
  \bibinfo{author}{\bibfnamefont{H.}~\bibnamefont{Berger}},
  \bibnamefont{et~al.}, \bibinfo{journal}{Phys. Rev. B}
  \textbf{\bibinfo{volume}{79}}, \bibinfo{pages}{045116}
  (\bibinfo{year}{2009}{\natexlab{a}}).

\bibitem[{\citenamefont{Cercellier et~al.}(2007)\citenamefont{Cercellier,
  Monney, Clerc, Battaglia, Despont, Garnier, Beck, Aebi, Patthey, Berger
  et~al.}}]{Cercellier}
\bibinfo{author}{\bibfnamefont{H.}~\bibnamefont{Cercellier}},
  \bibinfo{author}{\bibfnamefont{C.}~\bibnamefont{Monney}},
  \bibinfo{author}{\bibfnamefont{F.}~\bibnamefont{Clerc}},
  \bibinfo{author}{\bibfnamefont{C.}~\bibnamefont{Battaglia}},
  \bibinfo{author}{\bibfnamefont{L.}~\bibnamefont{Despont}},
  \bibinfo{author}{\bibfnamefont{M.~G.} \bibnamefont{Garnier}},
  \bibinfo{author}{\bibfnamefont{H.}~\bibnamefont{Beck}},
  \bibinfo{author}{\bibfnamefont{P.}~\bibnamefont{Aebi}},
  \bibinfo{author}{\bibfnamefont{L.}~\bibnamefont{Patthey}},
  \bibinfo{author}{\bibfnamefont{H.}~\bibnamefont{Berger}},
  \bibnamefont{et~al.}, \bibinfo{journal}{Phys. Rev. Lett.}
  \textbf{\bibinfo{volume}{99}}, \bibinfo{pages}{146403}
  (\bibinfo{year}{2007}).

\bibitem[{\citenamefont{Morosan et~al.}(2006)\citenamefont{Morosan, Zandbergen,
  Dennis, Bos, Onose, Klimczuk, Ramirez, Ong, and Cava}}]{Morosan}
\bibinfo{author}{\bibfnamefont{E.}~\bibnamefont{Morosan}},
  \bibinfo{author}{\bibfnamefont{H.~W.} \bibnamefont{Zandbergen}},
  \bibinfo{author}{\bibfnamefont{B.~S.} \bibnamefont{Dennis}},
  \bibinfo{author}{\bibfnamefont{J.~W.~G.} \bibnamefont{Bos}},
  \bibinfo{author}{\bibfnamefont{Y.}~\bibnamefont{Onose}},
  \bibinfo{author}{\bibfnamefont{T.}~\bibnamefont{Klimczuk}},
  \bibinfo{author}{\bibfnamefont{A.~P.} \bibnamefont{Ramirez}},
  \bibinfo{author}{\bibfnamefont{N.~P.} \bibnamefont{Ong}}, \bibnamefont{and}
  \bibinfo{author}{\bibfnamefont{R.~J.} \bibnamefont{Cava}},
  \bibinfo{journal}{Nat. Phys.} \textbf{\bibinfo{volume}{2}},
  \bibinfo{pages}{544} (\bibinfo{year}{2006}).

\bibitem[{\citenamefont{Morosan et~al.}(2007)\citenamefont{Morosan, Li, Ong,
  and Cava}}]{Morosan2}
\bibinfo{author}{\bibfnamefont{E.}~\bibnamefont{Morosan}},
  \bibinfo{author}{\bibfnamefont{L.}~\bibnamefont{Li}},
  \bibinfo{author}{\bibfnamefont{N.~P.} \bibnamefont{Ong}}, \bibnamefont{and}
  \bibinfo{author}{\bibfnamefont{R.~J.} \bibnamefont{Cava}},
  \bibinfo{journal}{Phys. Rev. B} \textbf{\bibinfo{volume}{75}},
  \bibinfo{pages}{104505} (\bibinfo{year}{2007}).

\bibitem[{\citenamefont{Kusmartseva et~al.}(2009)\citenamefont{Kusmartseva,
  Sipos, Berger, Forr{\'o}, and Tuti{\v s}}}]{Anna}
\bibinfo{author}{\bibfnamefont{A.~F.} \bibnamefont{Kusmartseva}},
  \bibinfo{author}{\bibfnamefont{B.}~\bibnamefont{Sipos}},
  \bibinfo{author}{\bibfnamefont{H.}~\bibnamefont{Berger}},
  \bibinfo{author}{\bibfnamefont{L.}~\bibnamefont{Forr{\'o}}},
  \bibnamefont{and} \bibinfo{author}{\bibfnamefont{E.}~\bibnamefont{Tuti{\v
  s}}}, \bibinfo{journal}{Phys. Rev. Lett.} \textbf{\bibinfo{volume}{103}},
  \bibinfo{pages}{236401} (\bibinfo{year}{2009}).

\bibitem[{\citenamefont{Klipstein and Friend}(1984)}]{Klipstein}
\bibinfo{author}{\bibfnamefont{P.~C.} \bibnamefont{Klipstein}}
  \bibnamefont{and} \bibinfo{author}{\bibfnamefont{R.~H.}
  \bibnamefont{Friend}}, \bibinfo{journal}{J. Phys. C}
  \textbf{\bibinfo{volume}{17}}, \bibinfo{pages}{2713} (\bibinfo{year}{1984}).

\bibitem[{\citenamefont{Friend and Yoffe}(1987)}]{Friend}
\bibinfo{author}{\bibfnamefont{R.~H.} \bibnamefont{Friend}} \bibnamefont{and}
  \bibinfo{author}{\bibfnamefont{A.~D.} \bibnamefont{Yoffe}},
  \bibinfo{journal}{Adv. Phys.} \textbf{\bibinfo{volume}{36}},
  \bibinfo{pages}{1} (\bibinfo{year}{1987}).

\bibitem[{\citenamefont{Li et~al.}(2007{\natexlab{b}})\citenamefont{Li, Wu,
  Chen, and Taillefer}}]{Li}
\bibinfo{author}{\bibfnamefont{S.~Y.} \bibnamefont{Li}},
  \bibinfo{author}{\bibfnamefont{G.}~\bibnamefont{Wu}},
  \bibinfo{author}{\bibfnamefont{X.~H.} \bibnamefont{Chen}}, \bibnamefont{and}
  \bibinfo{author}{\bibfnamefont{L.}~\bibnamefont{Taillefer}},
  \bibinfo{journal}{Phys. Rev. Lett.} \textbf{\bibinfo{volume}{99}},
  \bibinfo{pages}{107001} (\bibinfo{year}{2007}{\natexlab{b}}).

\bibitem[{\citenamefont{Jishi and Alyahyaei}(2008)}]{Jishi}
\bibinfo{author}{\bibfnamefont{R.~A.} \bibnamefont{Jishi}} \bibnamefont{and}
  \bibinfo{author}{\bibfnamefont{H.~M.} \bibnamefont{Alyahyaei}},
  \bibinfo{journal}{Phys. Rev. B} \textbf{\bibinfo{volume}{78}},
  \bibinfo{pages}{144516} (\bibinfo{year}{2008}).

\bibitem[{\citenamefont{van Wezel et~al.}(2010{\natexlab{c}})\citenamefont{van
  Wezel, Nahai-Williamson, and Saxena}}]{ourselvesDresden}
\bibinfo{author}{\bibfnamefont{J.}~\bibnamefont{van Wezel}},
  \bibinfo{author}{\bibfnamefont{P.}~\bibnamefont{Nahai-Williamson}},
  \bibnamefont{and} \bibinfo{author}{\bibfnamefont{S.~S.}
  \bibnamefont{Saxena}}, \bibinfo{journal}{Phys. Stat. Sol. b}
  \textbf{\bibinfo{volume}{247}}, \bibinfo{pages}{592}
  (\bibinfo{year}{2010}{\natexlab{c}}).

\bibitem[{\citenamefont{Sachdev}(2001)}]{Sachdev}
\bibinfo{author}{\bibfnamefont{S.}~\bibnamefont{Sachdev}},
  \emph{\bibinfo{title}{Quantum Phase Transitions.}}
  (\bibinfo{publisher}{Cambridge University Press}, \bibinfo{year}{2001}).

\bibitem[{\citenamefont{Moriya}(1985)}]{Moriya}
\bibinfo{author}{\bibfnamefont{T.}~\bibnamefont{Moriya}},
  \emph{\bibinfo{title}{Spin Fluctuations in Itinerant Electron Magnetism.}}
  (\bibinfo{publisher}{Springer, Berlin}, \bibinfo{year}{1985}).

\bibitem[{\citenamefont{Lonzarich}(1997)}]{Lonzarich}
\bibinfo{author}{\bibfnamefont{G.~G.} \bibnamefont{Lonzarich}}, in
  \emph{\bibinfo{booktitle}{Electron.}}, edited by
  \bibinfo{editor}{\bibfnamefont{M.}~\bibnamefont{Springford}}
  (\bibinfo{publisher}{Cambridge University Press}, \bibinfo{year}{1997}).

\bibitem[{\citenamefont{Lonzarich and Taillefer}(1985)}]{Lonzarich85}
\bibinfo{author}{\bibfnamefont{G.~G.} \bibnamefont{Lonzarich}}
  \bibnamefont{and}
  \bibinfo{author}{\bibfnamefont{L.}~\bibnamefont{Taillefer}},
  \bibinfo{journal}{J. Phys. C} \textbf{\bibinfo{volume}{18}},
  \bibinfo{pages}{4339} (\bibinfo{year}{1985}).

\bibitem[{\citenamefont{Saxena et~al.}(2000)\citenamefont{Saxena, Agarwal,
  Ahilan, Grosche, Haselwimmer, Steiner, Pugh, Walker, Julian, Monthoux
  et~al.}}]{Saxena}
\bibinfo{author}{\bibfnamefont{S.~S.} \bibnamefont{Saxena}},
  \bibinfo{author}{\bibfnamefont{P.}~\bibnamefont{Agarwal}},
  \bibinfo{author}{\bibfnamefont{K.}~\bibnamefont{Ahilan}},
  \bibinfo{author}{\bibfnamefont{F.~M.} \bibnamefont{Grosche}},
  \bibinfo{author}{\bibfnamefont{R.~K.~W.} \bibnamefont{Haselwimmer}},
  \bibinfo{author}{\bibfnamefont{M.~J.} \bibnamefont{Steiner}},
  \bibinfo{author}{\bibfnamefont{E.}~\bibnamefont{Pugh}},
  \bibinfo{author}{\bibfnamefont{I.~R.} \bibnamefont{Walker}},
  \bibinfo{author}{\bibfnamefont{S.~R.} \bibnamefont{Julian}},
  \bibinfo{author}{\bibfnamefont{P.}~\bibnamefont{Monthoux}},
  \bibnamefont{et~al.}, \bibinfo{journal}{Nature}
  \textbf{\bibinfo{volume}{406}}, \bibinfo{pages}{587} (\bibinfo{year}{2000}).

\bibitem[{\citenamefont{Lonzarich et~al.}(2008)\citenamefont{Lonzarich,
  Monthoux, and Pines}}]{Pines}
\bibinfo{author}{\bibfnamefont{G.~G.} \bibnamefont{Lonzarich}},
  \bibinfo{author}{\bibfnamefont{P.}~\bibnamefont{Monthoux}}, \bibnamefont{and}
  \bibinfo{author}{\bibfnamefont{D.}~\bibnamefont{Pines}},
  \bibinfo{journal}{Nature} \textbf{\bibinfo{volume}{450}},
  \bibinfo{pages}{1117} (\bibinfo{year}{2008}).

\bibitem[{\citenamefont{Smith et~al.}(2008)\citenamefont{Smith, Sutherland,
  Lonzarich, Saxena, Kimura, Takashima, Nohara, and Takagi}}]{Smith}
\bibinfo{author}{\bibfnamefont{R.~P.} \bibnamefont{Smith}},
  \bibinfo{author}{\bibfnamefont{M.}~\bibnamefont{Sutherland}},
  \bibinfo{author}{\bibfnamefont{G.~G.} \bibnamefont{Lonzarich}},
  \bibinfo{author}{\bibfnamefont{S.~S.} \bibnamefont{Saxena}},
  \bibinfo{author}{\bibfnamefont{N.}~\bibnamefont{Kimura}},
  \bibinfo{author}{\bibfnamefont{S.}~\bibnamefont{Takashima}},
  \bibinfo{author}{\bibfnamefont{M.}~\bibnamefont{Nohara}}, \bibnamefont{and}
  \bibinfo{author}{\bibfnamefont{H.}~\bibnamefont{Takagi}},
  \bibinfo{journal}{Nature} \textbf{\bibinfo{volume}{455}},
  \bibinfo{pages}{1220} (\bibinfo{year}{2008}).

\bibitem[{\citenamefont{Rowley et~al.}(2009)\citenamefont{Rowley, Spalek,
  Smith, Dean, Lonzarich, Scott, and Saxena}}]{Rowley}
\bibinfo{author}{\bibfnamefont{S.~E.} \bibnamefont{Rowley}},
  \bibinfo{author}{\bibfnamefont{L.~J.} \bibnamefont{Spalek}},
  \bibinfo{author}{\bibfnamefont{R.~P.} \bibnamefont{Smith}},
  \bibinfo{author}{\bibfnamefont{M.~P.~M.} \bibnamefont{Dean}},
  \bibinfo{author}{\bibfnamefont{G.~G.} \bibnamefont{Lonzarich}},
  \bibinfo{author}{\bibfnamefont{J.~F.} \bibnamefont{Scott}}, \bibnamefont{and}
  \bibinfo{author}{\bibfnamefont{S.~S.} \bibnamefont{Saxena}}
  (\bibinfo{year}{2009}), \bibinfo{note}{arxiv:Cond-mat 0903.1445}.

%
\bibitem[{\citenamefont{Monney et~al.}(2009{\natexlab{b}})\citenamefont{Monney,
  Schwier, Garnier, Battaglia, Mariotti, Didiot, Cercellier, Marcus, Berger,
  Titov et~al.}}]{MonneyPLD}
\bibinfo{author}{\bibfnamefont{C.}~\bibnamefont{Monney}},
  \bibinfo{author}{\bibfnamefont{E.~F.} \bibnamefont{Schwier}},
  \bibinfo{author}{\bibfnamefont{M.~G.} \bibnamefont{Garnier}},
  \bibinfo{author}{\bibfnamefont{C.}~\bibnamefont{Battaglia}},
  \bibinfo{author}{\bibfnamefont{N.}~\bibnamefont{Mariotti}},
  \bibinfo{author}{\bibfnamefont{C.}~\bibnamefont{Didiot}},
  \bibinfo{author}{\bibfnamefont{H.}~\bibnamefont{Cercellier}},
  \bibinfo{author}{\bibfnamefont{J.}~\bibnamefont{Marcus}},
  \bibinfo{author}{\bibfnamefont{H.}~\bibnamefont{Berger}},
  \bibinfo{author}{\bibfnamefont{A.~N.} \bibnamefont{Titov}},
  \bibnamefont{et~al.}, \bibinfo{journal}{ArXiv:cond-mat} p.
  \bibinfo{pages}{0912.5283v1} (\bibinfo{year}{2009}{\natexlab{b}}).

\bibitem[{\citenamefont{Varma et~al.}(1987)\citenamefont{Varma, Schmitt-Rink,
  and Abrahams}}]{Varma}
\bibinfo{author}{\bibfnamefont{C.~M.} \bibnamefont{Varma}},
  \bibinfo{author}{\bibfnamefont{S.}~\bibnamefont{Schmitt-Rink}},
  \bibnamefont{and} \bibinfo{author}{\bibfnamefont{E.}~\bibnamefont{Abrahams}},
  \bibinfo{journal}{Solid State Comm.} \textbf{\bibinfo{volume}{62}},
  \bibinfo{pages}{681} (\bibinfo{year}{1987}).

\bibitem[{\citenamefont{Littlewood}(1990)}]{Littlewood}
\bibinfo{author}{\bibfnamefont{P.~B.} \bibnamefont{Littlewood}},
  \bibinfo{journal}{Phys. Rev. B} \textbf{\bibinfo{volume}{42}},
  \bibinfo{pages}{10075} (\bibinfo{year}{1990}).

\bibitem[{\citenamefont{Nakai et~al.}(2002)\citenamefont{Nakai, Ichioka, and
  Machida}}]{Nakai}
\bibinfo{author}{\bibfnamefont{N.}~\bibnamefont{Nakai}},
  \bibinfo{author}{\bibfnamefont{M.}~\bibnamefont{Ichioka}}, \bibnamefont{and}
  \bibinfo{author}{\bibfnamefont{K.}~\bibnamefont{Machida}},
  \bibinfo{journal}{J. Phys. Soc. Japan} \textbf{\bibinfo{volume}{71}},
  \bibinfo{pages}{23} (\bibinfo{year}{2002}).

\bibitem[{\citenamefont{Kiss et~al.}(2007)\citenamefont{Kiss, Yokoya, Chainani,
  Shin, Hanaguri, Nohara, and Takagi}}]{Kiss}
\bibinfo{author}{\bibfnamefont{T.}~\bibnamefont{Kiss}},
  \bibinfo{author}{\bibfnamefont{T.}~\bibnamefont{Yokoya}},
  \bibinfo{author}{\bibfnamefont{A.}~\bibnamefont{Chainani}},
  \bibinfo{author}{\bibfnamefont{S.}~\bibnamefont{Shin}},
  \bibinfo{author}{\bibfnamefont{T.}~\bibnamefont{Hanaguri}},
  \bibinfo{author}{\bibfnamefont{M.}~\bibnamefont{Nohara}}, \bibnamefont{and}
  \bibinfo{author}{\bibfnamefont{H.}~\bibnamefont{Takagi}},
  \bibinfo{journal}{Nat. Phys.} \textbf{\bibinfo{volume}{3}},
  \bibinfo{pages}{720} (\bibinfo{year}{2007}).

\bibitem[{\citenamefont{Fr\"ohlich et~al.}(1950)\citenamefont{Fr\"ohlich,
  Pelzer, and Zienau}}]{Frohlich1}
\bibinfo{author}{\bibfnamefont{H.}~\bibnamefont{Fr\"ohlich}},
  \bibinfo{author}{\bibfnamefont{H.}~\bibnamefont{Pelzer}}, \bibnamefont{and}
  \bibinfo{author}{\bibfnamefont{S.}~\bibnamefont{Zienau}},
  \bibinfo{journal}{Phil. Mag.} \textbf{\bibinfo{volume}{41}},
  \bibinfo{pages}{221} (\bibinfo{year}{1950}).

\bibitem[{\citenamefont{Fr\"ohlich}(1954)}]{Frohlich2}
\bibinfo{author}{\bibfnamefont{H.}~\bibnamefont{Fr\"ohlich}},
  \bibinfo{journal}{Adv. Phys.} \textbf{\bibinfo{volume}{3}},
  \bibinfo{pages}{325} (\bibinfo{year}{1954}).

\bibitem[{\citenamefont{Kuper and Whitfield}(1963)}]{Frohlich3}
\bibinfo{editor}{\bibfnamefont{C.~G.} \bibnamefont{Kuper}} \bibnamefont{and}
  \bibinfo{editor}{\bibfnamefont{G.~D.} \bibnamefont{Whitfield}}, eds.,
  \emph{\bibinfo{title}{Polarons and excitons : Scottish Universities' Summer
  School, 1962}} (\bibinfo{publisher}{Oliver and Boyd, Edinburgh},
  \bibinfo{year}{1963}).

\bibitem[{\citenamefont{Ziman}(1960)}]{Ziman}
\bibinfo{author}{\bibfnamefont{J.}~\bibnamefont{Ziman}},
  \emph{\bibinfo{title}{Electrons and Phonons.}}
  (\bibinfo{publisher}{Clarendon, Oxford}, \bibinfo{year}{1960}).

\bibitem[{\citenamefont{Dungate}(1990)}]{Dungate}
\bibinfo{author}{\bibfnamefont{D.~G.} \bibnamefont{Dungate}}, Ph.D. thesis,
  \bibinfo{school}{University of Cambridge} (\bibinfo{year}{1990}),
  \bibinfo{note}{phd.17451}.

\bibitem[{\citenamefont{Hlubina and Rice}(1995)}]{Rice}
\bibinfo{author}{\bibfnamefont{R.}~\bibnamefont{Hlubina}} \bibnamefont{and}
  \bibinfo{author}{\bibfnamefont{T.~M.} \bibnamefont{Rice}},
  \bibinfo{journal}{Phys. Rev. B} \textbf{\bibinfo{volume}{51}},
  \bibinfo{pages}{9253} (\bibinfo{year}{1995}).

\bibitem[{\citenamefont{Ueda}(1977)}]{Ueda}
\bibinfo{author}{\bibfnamefont{K.}~\bibnamefont{Ueda}}, \bibinfo{journal}{J.
  Phys. Soc. Jpn.} \textbf{\bibinfo{volume}{43}}, \bibinfo{pages}{1497}
  (\bibinfo{year}{1977}).

\bibitem[{\citenamefont{{See for exmaple: A. H. Wilson}}(1953)}]{WilsonBook}
\bibinfo{author}{\bibnamefont{{See for exmaple: A. H. Wilson}}},
  \emph{\bibinfo{title}{The Theory of Metals.}} (\bibinfo{publisher}{Cambridge
  University Press, Cambridge}, \bibinfo{year}{1953}).

\bibitem[{\citenamefont{Wilson}(1938)}]{Wilson2}
\bibinfo{author}{\bibfnamefont{A.~H.} \bibnamefont{Wilson}},
  \bibinfo{journal}{Proc. R. Soc. Lond. A} \textbf{\bibinfo{volume}{167}},
  \bibinfo{pages}{580} (\bibinfo{year}{1938}).

\bibitem[{\citenamefont{Ueda and Moriya}(1975)}]{Ueda2}
\bibinfo{author}{\bibfnamefont{K.}~\bibnamefont{Ueda}} \bibnamefont{and}
  \bibinfo{author}{\bibfnamefont{T.}~\bibnamefont{Moriya}},
  \bibinfo{journal}{J. Phys. Soc. Jpn.} \textbf{\bibinfo{volume}{39}},
  \bibinfo{pages}{605} (\bibinfo{year}{1975}).

\end{thebibliography}

\end{document}